\documentclass[achemso,showpacs,floatfix,superscriptaddress,twocolumn]{revtex4-1}
\usepackage{graphicx}    
\usepackage{hyperref}    
\usepackage{bm}          	
\usepackage[sumlimits,intlimits]{amsmath}
\usepackage{amsfonts,amssymb}
\usepackage[section]{placeins}

\begin{document}

\title{Valleytronics in Tin (II) Sulfide}

\author{A. S. Rodin}
\author{L. C. Gomes}
\author{A. Carvalho}
\author{A. H. Castro Neto}

\affiliation{Centre for Advanced 2D Materials and Graphene Research Centre,
 National University of Singapore, 6 Science Drive 2, 117546, Singapore}

\date{\today}
\begin{abstract}

Tin (II) sulfide (SnS) is a layered mineral found in nature. In this paper, we study the two-dimensional form of this material using a combination of \emph{ab initio} calculation and $\mathbf{k}\cdot\mathbf{p}$ theory. In particular, we address the valley properties and the optical selection rules of 2D SnS. Our study reveals SnS as an extraordinary material for valleytronics, 
where pairs of equivalent valleys are placed along two perpendicular axes,  can be selected exclusively with linear polarized light,
and can be separated using non-local electrical measurements.

\end{abstract}


\maketitle

\emph{Introduction.} The ever-expanding library of two-dimensional (2D) materials can be generally divided into several classes. One of them is composed of mono-elemental structures: the pioneering graphene and, as more recent additions, silicene and phosphorene. While graphene is the best known member of this group, phosphorene has been experiencing a rapid growth in the scientific community due to a number of factors, including its mobility, high anisotropy, and gap tunability by strain and sample thickness.\cite{rodin,liu,low,zhang,avsar}

Another important class contains the transition metal dichalcogenides (TMDC's), such as $\text{MoSe}_2$, $\text{MoS}_2$, $\text{WSe}_2$, and $\text{WS}_2$, among others.\cite{goki} These materials are semiconductors with different gap sizes, a substantial spin-orbit coupling, and a valley degree of freedom\cite{zeng-NN-7-490,mak-NN-7-494,cao-NC-3-887}. These properties make TMDC's attractive for potential spintronic\cite{wolf} and valleytronic\cite{rycerz-NP-3-172,xiao-PRL-99-236809} applications.

Recently, a new type of layered structures, referred to as group-IV monochalcogenides, has been brought to the attention of the physics community\cite{gomes}. As the name suggests, these materials are composed of equal parts of group-IV metals and chalcogen atoms, and have a general chemical formula MC. In earlier works\cite{tritsaris,vidal,brad,singh-APL-105-042103,gomes}, several of these materials have been investigated to determine their structural, electronic, and optical properties. These novel materials combine a number of features that make them very attractive for future study. Like TMDC's, monochalcogenides also possess multiple valleys. However, because of the rectangular unit cell, the valleys are situated on the axes of the Brillouin zone. Because of this, valley separation can be performed with linearly polarized light instead of the circular polarization required for TMDC's. In addition, perpendicular orientation of the valleys makes is possible to separate the valleys using the transverse non-linear conductivity. Finally, the presence of heavy elements leads to a significant spin-orbit coupling. Thus, these materials are excellent canditates for optical, valleytronic, and spintronic applications.

In this paper, we focus on one of the monochalcogenides: tin (II) sulfide (SnS). The rationale for our choice is the fact that SnS is found in nature in a mineral form, known as herzenbergite. To study the electronic properties of this material, we use \emph{ab initio} calculations in conjunction with standard numerical and analytical methods. We develop a $\mathbf{k}\cdot\mathbf{ p}$ model to describe the band structure at different points in the Brillouin zone and use these results to analyze the optical selection rules, as well as the nonlinear transverse conductivity due to the valleys. Lastly, we discuss how those properties can be used to read and write the valley quantum number, laying a foundation for the use of SnS as a functional material for valleytronics.

\begin{figure}[h]
\includegraphics[width = 2.5in]{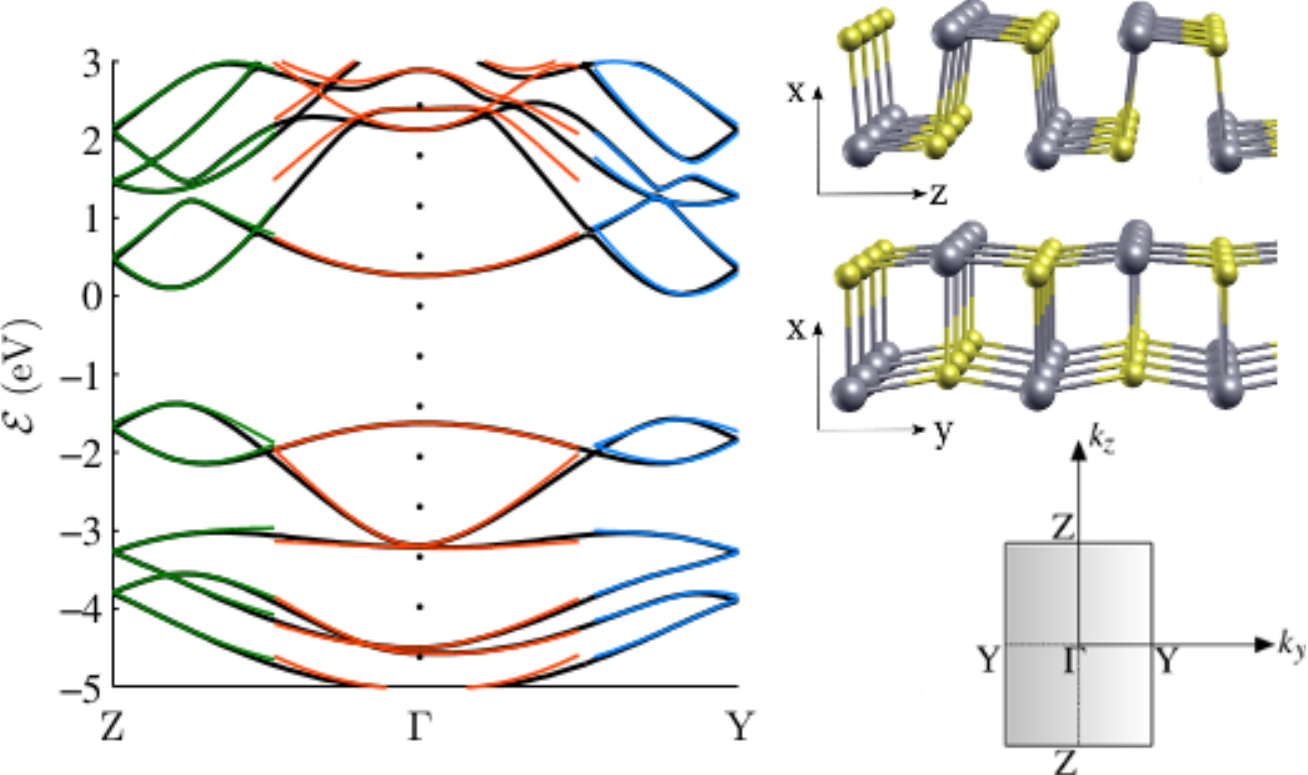}
\caption{(color online) $\mathbf{k}\cdot\mathbf{p}$ expansions around the $\Gamma$ point and the valley extrema (color lines) plotted against the \emph{ab initio} band structure (black lines). The crystal lattice of a monolayer SnS in two different projections and the Brillouin zone are also shown.}
\label{fig:k_dot_p}
\end{figure}

\emph{Electronic Structure.} Monochalcogenides of group-IV elements are isoelectronic with phosphorus, and their structure is reminiscent of corrugated phosphorene. However, because of the two atomic species, the symmetry of the crystal is lowered and bulk SnS belongs to the Pnma-$\text{D}^{16}_{2h}$ space group. Going from bulk to monolayer, SnS loses  the inversion symmetry, putting it in the Pmn$2_1$ space group. There are four symmetry transformations in this group: the identity, a reflection, a reflection with a translation, and a rotation with a translation. The irreducible representations (irreps) for this space group, along with the operators belonging to each irrep are listed in Table~\ref{Tab:Irreps}. Following the standard convention, the axis of highest symmetry is labeled as $z$ and it runs perpendicular to the corrugations. The $y$-axis points along the corrugations; $x$ axis is normal to the plane of the monolayer, see Fig.~\ref{fig:k_dot_p}. 
\begin{table}[h]
\begin{ruledtabular}
\begin{tabular}{lccccc}
 & $E$ & $R_y$ &  $\tau R_x$& $\tau C_z$&\\
\hline
$A_1$   &  1    & 1 & 1  & 1 &$z$\\ 
\hline
$B_1$   &  1    & 1 & -1 & -1  &$x$, $L_y$ \\
\hline
$B_2$   &  1   & -1 & 1 & -1 & $y$, $L_x$\\
\hline
$A_2$   &  1   & -1&-1&1 &$L_z$
\end{tabular}
\end{ruledtabular}
\caption{
Irreducible representations and their transformations under symmetry operations for the  Pmn$2_1$ space group. $R_i$ represents reflection across the plane normal to the $i$ axis; $C_i$ is the two-fold rotation around the $i$ axis. The $\tau$ before the operation means that the lattice is shifted by $(0,\,a_y/2,\,a_z/2)$ after the operation.}
\label{Tab:Irreps}
\end{table}
The direct products of the Pmn$2_1$ irreps are listed in Table~\ref{Tab:Products}.
\begin{table}[h]
\begin{ruledtabular}
\begin{tabular}{l@{\ \ \ }|cccc}
$\otimes$  & $A_1$ & $B_1$ &  $B_2$& $A_2$\\
\hline
$A_1$   &  $A_1$    &  $B_1$ &  $B_2$  & $A_2$\\ 
$B_1$   &  $B_1$    &  $A_1$ &  $A_2$  & $B_2$ \\
$B_2$  &  $B_2$    &  $A_2$ &  $A_1$  & $B_1$ \\
$A_2$  &  $A_2$    &  $B_2$ &  $B_1$  & $A_1$
\end{tabular}
\end{ruledtabular}
\caption{
Direct products of the irreducible representations for the  Pmn$2_1$ space group.}
\label{Tab:Products}
\end{table}

The band structure of monochancogenides is distinct from both TMDC's and phosphorene and contains two pairs of valleys, located on the $Z\Gamma$ and $Y\Gamma$ axes. We obtain a few-band Hamiltonian at these valleys using the $\mathbf{k}\cdot\mathbf{p}$ formalism.

The $\mathbf{k}\cdot\mathbf{p}$ Hamiltonian is constructed by finding the exact eigenstates and the corresponding eigenvalues at a point $\mathbf{k}_0$ in the Brillouin zone, described by the Hamiltonian $H_0$. One then treats the deviation from $\mathbf{k}_0$ as a perturbation, given by
\begin{equation}
H_1 = \left[\frac{\hbar^2k^2}{2m}+\frac{\hbar^2}{m}\left(\hat{\mathbf{p}}+\mathbf{k_0}\right)\cdot \mathbf{k}\right]\,,
\label{eqn:H_1}
\end{equation}
where $H_0+H_1 = H_\text{full}$, $m$ is the bare electron mass, $\hat{\mathbf{p}}$ is the momentum operator, and $\mathbf{k}$ is measured from the position $\mathbf{k_0}$.

We begin our analysis of the band structure by plotting the \emph{ab initio} results along with the $\mathbf{k}\cdot\mathbf{p}$ approximation in Fig.~\ref{fig:k_dot_p}. 
The first-principles calculations were performed using the {\sc Quantum ESPRESSO} code.\cite{Giannozzi2009}. 
The exchange correlation energy was described by the generalized
gradient approximation (GGA) using the PBE\cite{pbe} functional.
Interactions between valence electrons and ionic cores are described by Troullier-Martins pseudopotentials\cite{tm}.
The Kohn-Sham orbitals were expanded in a plane-wave basis with a cutoff energy of 65~Ry, and for the charge density, a cutoff of
280~Ry was used. 
For the self-consistent calculation of the charge density, the Brillouin-zone (BZ) was sampled using a $\Gamma$-centered 1$\times$10$\times$10 grid following the scheme
proposed by Monkhorst-Pack\cite{Monkhorst1968}. For the calculation of the dipole matrix elements, a finer
1$\times$40$\times$40 grid was employed.

Three independent $\mathbf{k}\cdot\mathbf{p}$ expansions are performed---one is around the $\Gamma$ point, the other two are around the valley extrema. For this, we used 30 energy states with the corresponding matrix elements of $H_1$ obtained from the density functional theory calculations.

From Fig.~\ref{fig:k_dot_p}, it is clear that the $\mathbf{k}\cdot\mathbf{p}$ approximation is quite good at capturing the bands. Of course, using 30 levels is impractical. Therefore, we construct smaller Hamiltonians around the centers of the valleys by judiciously choosing which bands can be dropped, while still describing the conduction and the valence bands correctly.

To obtain the $\mathbf{k}\cdot\mathbf{p}$ Hamiltonian, we calculate the matrix elements of $H_1$ for the known eigenstates at $\mathbf{k}_0$, given by $ \langle i|H_1|j\rangle$, where $|i\rangle$ and $|j\rangle$ are the eigenstates. It is clear from Eq.~\eqref{eqn:H_1} that the terms proportional to $k^2$ and $\mathbf{k}_0\cdot\mathbf{k}$  are scalar and, thus, finite only if $|i\rangle=|j\rangle$. On the other hand, the $\hat{\mathbf{p}}\cdot\mathbf{k} = \hat{p}_zk_z+\hat{p}_yk_y$ term can couple different bands. One can determine which matrix elements are finite based on the symmetries of the individual states.

For a non-vanishing matrix element, the integral over all space of $\langle i|\hat p_n|j\rangle$ has to be nonzero. This means that the term inside the integral must not be anti-symmetric for any of the transformations of the space group. In other words, the product must transform as the $A_1$ irrep, see Table~\ref{Tab:Irreps}.

\begin{figure}[h]
\includegraphics[width = 3in]{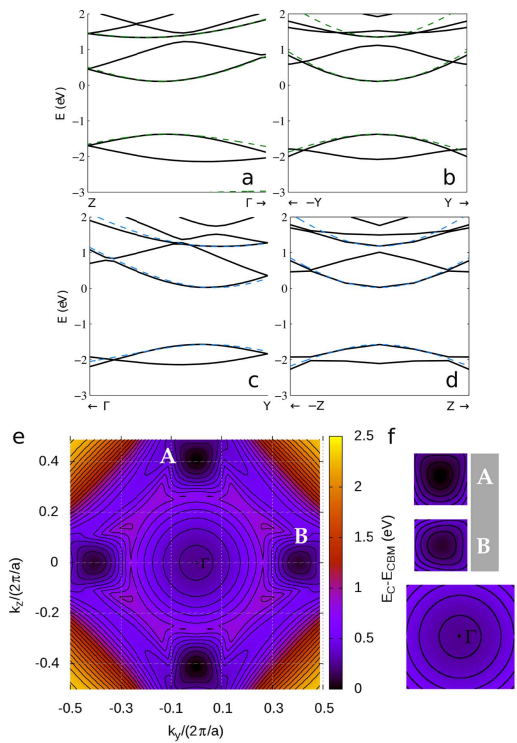}
\caption{Cross-sections of $\mathbf{k}\cdot\mathbf{p}$ results for a $4\times 4$ Hamiltonian for (a,b) the $Z\Gamma$ and (c,d) the $Y\Gamma$ valleys. The dashed lines are the $\mathbf{k}\cdot\mathbf{p}$ results, while the solid lines are obtained from \emph{ab initio} calculations.
(e)  Conduction band profile as obtained from first principles calculations and (f) $\mathbf{k}\cdot\mathbf{p}$ 
conduction band profile at the minima, obtained from the $\mathbf{k}\cdot\mathbf{p}$ results for a $4\times 4$ Hamiltonian.}
\label{fig:VBM}
\end{figure}

Let us first address the valley located along the $Z\Gamma$ line. From the first principles, we determine that both the conduction and the valence bands belong to the $A_1$ irrep. Starting with the matrix element for $\hat p_z$, which is a part of the $A_1$ irrep, we see that the direct product of the irreps for the bands and the operator is given by $A_1\otimes A_1\otimes A_1 = A_1$, according to Table~\ref{Tab:Products}, leading to a finite matrix element for $\hat p_z$. When we attempt to do the same for $\hat p_y$ ($B_2$ irrep), we get $A_1\otimes B_2\otimes A_1 = B_2\neq A_1$, resulting in a vanishing matrix element. In fact, both the valence and the conduction band couple only to $B_2$ states in $y$ direction ($A_1\otimes B_2\otimes B_2 = A_1$). Thus, to ensure a correct dispersion in the $y$ direction, we introduce more bands. It turns out that just two additional bands (the third valence and the third conduction bands, counting from those closest to the gap) are sufficient to capture the shape of the relevant valence and conduction bands, see Fig.~\ref{fig:VBM}(a,b). The Hamiltonian at the $Z\Gamma$ valley takes the following form
\begin{align}
H_{Z\Gamma} &=\tensor{\mathcal{E}}_{Z\Gamma}+\frac{\hbar^2k^2}{2m}\times\mathbf{1}+ \begin{pmatrix}
\lambda_{3v}&0&0&\gamma_2
\\
0&\lambda_{v}&\gamma_1&0
\\
0&\gamma_1^*&\lambda_c&0
\\
\gamma_2^*&0&0&\lambda_{3c}
\end{pmatrix}k_z+
\nonumber
\\
+ &\begin{pmatrix}
0&\beta_1&\beta_2&0
\\
\beta_1^*&0&0&\beta_3
\\
\beta_2^*&0&0&\beta_4
\\
0&\beta_3^*&\beta_4^*&0
\end{pmatrix}k_y\,,
\label{eqn:VBM}
\end{align}
where $\tensor{\mathcal{E}}_{Z\Gamma}$ is a matrix with the band energies $(E_{3v}, E_v,E_c,E_{3c})$ on the diagonal. The values of other matrix elements are provided in the Supplementary Information.

Next, we move to the other pair of valleys, located on the $Y\Gamma$ line. Because the point of expansion is not on the high-symmetry $z$ axis, the bands are no longer described by the four irreps listed above. Instead, the states fall into two categories A' and A'', depending on their symmetry across the $yz$ plane. The coupling elements are finite only for the states in the same irrep, which is true for the valence and the conduction band. However, to obtain the correct dispersion, we need more bands. Just as before, the third valence and conduction bands are enough. The results for the $Y\Gamma$ valley are plotted in Fig.~\ref{fig:VBM}(c,d). Here, the Hamiltonian is
\begin{align}
H_{Y\Gamma} &=\tensor{\mathcal{E}}_{Y\Gamma}+\frac{\hbar^2k^2}{2m}\times\mathbf{1}+ \begin{pmatrix}
0&\kappa_1&\kappa_2&\kappa_3
\\
\kappa_1^*&0&\kappa_4&\kappa_5
\\
\kappa_2^*&\kappa_4^*&0&\kappa_6
\\
\kappa_3^*&\kappa_5^*&\kappa_6^*&0
\end{pmatrix}k_z+
\nonumber
\\
+ &\begin{pmatrix}
\nu_{3v}&\phi_1&\phi_2&\phi_3
\\
\phi_1^*&\nu_{v}&\phi_4&\phi_5
\\
\phi_2^*&\phi_4^*&\nu_{c}&\phi_6
\\
\phi_3^*&\phi_5^*&\phi_6^*&\nu_{3c}
\end{pmatrix}k_y\,.
\label{eqn:CBM}
\end{align}
The values of the interband coupling elements are also given in the Supplementary Information.

\begin{figure}[h]
\includegraphics[width = 2.5in]{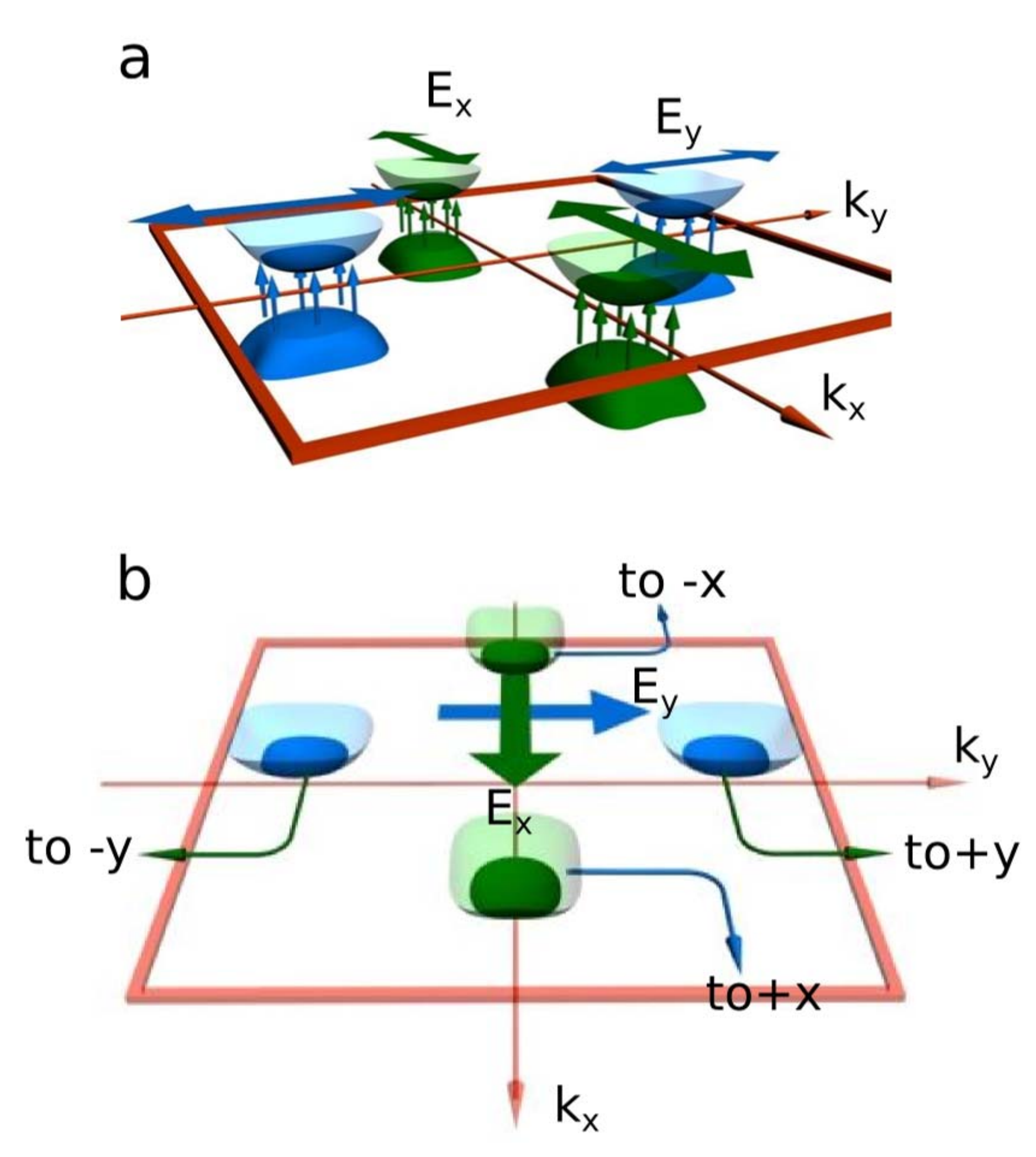}
\caption{Valley selection and separation. (a) Valley selection for an external oscillating electric field. Depending on the polarization of the field, different valleys are excited. (b) Valley separation under a static electric field. Depending on the polarization of the field, different valleys flow in the perpendicular direction.}
\label{fig:Valley_Selection}
\end{figure}

\emph{Valley separation.} Having determined the effective Hamiltonians for the two valleys, we are now in a position to discuss the possibility of employing the valley degree of freedom. To do so, we look at the transitions between the tops of the valence and the bottoms of the conduction bands for the two valley pairs. The coupling strength with the light polarized in $i$ direction is given by $\mathcal{C}_i = \frac{m}{\hbar}\langle c(\mathbf{k})| \frac{\partial H}{\partial k_i}|v(\mathbf{k})\rangle$. In other words, at $\mathbf{k}\rightarrow 0$, the coupling strength is proportional to the off-diagonal coefficient of the linear momentum term pointing in the direction of the light polarization. The transition rate is proportional to the coupling strength squared. As we established earlier, there is no coupling between the conduction and the valence bands at the $Z\Gamma$ valley in $y$ direction. This means that only light polarized in $z$ direction can induce transitions between these two bands. At the $Y\Gamma$ valley, on the other hand, both $y$ and $z$ polarized light can excite electrons. However, from the \emph{ab inition} calculation, we see that the transition rate for $y$ polarized light at this valley is approximately equal to the transition rate for $z$ polarized light at the $Z\Gamma$. Also, both almost 40  times larger than the transition rate for the $z$-polarized light at the $Y\Gamma$, see Supplementary Information for the matrix elements. This allows one to use linearly polarized light to selectively excite electrons from the valence to the conduction bands for individual valleys, see Fig.~\ref{fig:Valley_Selection}(a). A similar finding was previously reported for TMDC's, where the authors proposed using circularly polarized light for selective valley excitation.\cite{zeng-NN-7-490,mak-NN-7-494,cao-NC-3-887}

\emph{Transverse Current.} The final topic that we address is the nonlinear response to the electric field in the context of multiple valleys. It can be shown that at zero temperature, the second order response to an applied electric field results in the following current
\begin{align}
\mathbf{j}^{(2)}(\mathbf{E}) &= -\frac{e}{2\hbar}\left(\frac{e\tau}{2\pi\hbar}\right)^2\oint_\mathcal{C} d\mathbf{q}\, \nabla_\mathbf{q}\left(\mathbf{E}\cdot\nabla_\mathbf{q}\mathcal{E}_\mathbf{q}\right)^2\,,
\end{align}
where $\tau$ is the electron scattering time. This is a standard result, but we derivation can be found in the Supplementary Information for the reader's convenience. If the electric field points along one of the crystal axes ($ \mathbf{ \hat i}$), the transverse component of the integrand (pointing in $\mathbf{\hat{j}}$ direction)  becomes
\begin{equation}
\nabla_\mathbf{q}\left(\mathbf{E}\cdot\nabla_\mathbf{q}\mathcal{E}_\mathbf{q}\right)^2=E_i^2\frac{\partial}{\partial q_j}\left(\frac{\partial \mathcal{E}_\mathbf{q}}{\partial q_i}\right)^2\,.
\end{equation}
The current is nonzero only if the valley does not have a mirror symmetry in the $\mathbf{\hat j}$ direction so the derivative with respect to $q_j$ does not give equal and opposite contribution on the two sides of the valley. Using this reasoning, one can clearly see from the contour plot in Fig.~\ref{fig:VBM} that the transverse current is significant only for the pair of valleys perpendicular to the field. That is, if the field points in $y$ direction, the transverse $z$ current will only be observed for the $Z\Gamma$ valleys. If, on the other hand, the field points in $z$ direction, the transverse current will only arise in the $Y\Gamma$ valleys.

Due to the valley structure, the transverse current in the individual valleys of every valley pair ($Z\Gamma$ and $Y\Gamma$) points in the opposite direction, so that the total valley current is zero. However, in a finite system, different valley components accumulate on different sides of the sample (see Fig.~\ref{fig:Valley_Selection}(b) for the current direction). 
This effect is similar to spin Hall effect, but instead of spin separation, here one has a separation of valleys.
Thus, the valley population can be read through the application of an electrical current and measuring the transverse valley current  using a valley filter as described in Ref.\onlinecite{xiao-PRL-99-236809}.
A similar device scheme can also be used to prepare a current with a defined valley state.

The analysis in this letter shows that 
the peculiar bandstructure of monolayer SnS and the rectangular shape of the underlying Brillouin Zone
result in the presence of the two pairs of valleys both in the conduction and valence band.
Each pair of valleys can be optically pumped separately by excitation with linearly polarized light,
thus writing the valley state.
The valley state can be read using the reverse process, or alternatively by
using the non-linear transverse valley conductivity arising from the off-centricity and anisotropy of the valleys.
This system thus allows for writing and reading without the need of circularly polarized light, or magnetic field,
thus presenting clear advantages over other semiconductor materials for spintronics and valleytronics.

The authors acknowledge the National Research Foundation, Prime Minister Office, Singapore, under its Medium Sized Centre  Programme and CRP award ”Novel 2D materials with tailored properties: beyond graphene” (R-144-000-295- 281).

\end{document}